\documentclass[aps,prl,superscriptaddress,showkeys,twocolumn,longbibliography]{revtex4-2}

\usepackage{bm,amssymb,amsmath,amsthm,mathtools,xcolor,array}

\definecolor{darkgreen}{rgb}{0.0,0.5,0.0}
\definecolor{darkpurple}{rgb}{0.5,0.0,0.5}

\newcommand{\derpar}[2] {\frac{\partial #1}{\partial #2}}
\newcommand{\HAT}[1] {\hat{\mathbf{#1}}}
\newcommand{\VEC}[1] {\mathbf{#1}}
 
 \newcommand{\Dr}{D_\text{r}}
\DeclareMathOperator{\Real}{Re}
 
\begin{document}
\title{Kinetic theory of motility induced phase separation for active Brownian particles}

\author{Rodrigo Soto}
\affiliation{Departamento de F\'{\i}sica, FCFM, Universidad de Chile, Santiago, Chile.}

\author{Martín Pinto}
\affiliation{Departamento de F\'{\i}sica, FCFM, Universidad de Chile, Santiago, Chile.}

\author{Ricardo Brito}
\affiliation{Departamento de Estructura de la Materia, Física Térmica y Electr\'onica and GISC, Universidad Complutense de Madrid, Spain.}

%\date{\today}
\begin{abstract} %<= 600 characters
When two active Brownian particles collide, they slide along each other until they can continue their free motion. For persistence lengths much larger than the particle diameter, the directors do not change, but the collision can be modeled as producing a net displacement on the particles compared to their free motion in the absence of the encounter. With these elements, a Boltzmann--Enskog-like kinetic theory is built. A linear stability analysis of the homogeneous state predicts a density instability resulting from the effective velocity reduction of tagged particles predicted by the theory.
\end{abstract}

\maketitle

\textit{Introduction.}
Non-inertial self-propelled active particles, even when the dominant interaction between particles is  purely repulsive, have a natural tendency to cluster~\cite{zhang2010collective,thompson2011lattice,fily2012athermal,redner2013structure,theurkauff2012dynamic,palacci2013living,buttinoni2013dynamical,levis2014clustering}. Qualitatively, the mechanism is quite simple. When two particles meet, they remain in contact for a finite time, during which they point in roughly constant directions. If the particle density $\rho$ is high enough, during the time they remain in contact and before one of them changes its direction and escapes, more particles can arrive, becoming the seed of a cluster. That is, clustering is a direct consequence of persistence in motion and excluded volume. 
Active Brownian particles (ABP)~\cite{romanczuk2012active,cates2013active,solon2015active}
is an ideal model to test this hypothesis. Here, particles move persistently  at constant speed $V$ along  directors that change slowly by rotational diffusion with coefficient $\Dr$, and  interact only by excluded volume. The equations of motion for the position $\VEC r_i$ and the director $\HAT n_i$ of the particle $i$ are
\begin{align}
\dot{\VEC{r}}_i &= V \HAT{n}_i+\VEC{F}_i, &\dot{\HAT{n}}_i &= \sqrt{2\Dr}\,\bm{\xi}_i(t)\times\HAT{n}_i,
\end{align}
where $\bm{\xi}_i$ are uncorrelated white noises and $\VEC F_i$ the hard-core interparticle force acting on $i$.
For the case of spherical ABP with diameter $\sigma$, despite the absence of interparticle attraction, persistence was indeed found to induce clustering~\cite{fily2012athermal,redner2013structure,levis2014clustering}. 
Dimensional analysis and simulations indicate that the relevant control parameters in $d$ spatial dimensions are $\tilde\rho=\rho\sigma^d$ (which is proportional to the volume fraction) and the persistence length $\ell=V/(\sigma \Dr)$, also called the active P\'eclet number. Clustering takes place for high $\tilde\rho$ and $\ell$~\cite{redner2013structure,levis2014clustering,fily2014freezing,digregorio2018full,van2019interparticle}.

A theoretical framework to describe the clustering instability is the so-called motility induced phase separation (MIPS), which states that the effective particle velocities are reduced due to particle encounters, which turn out to be a decreasing function of the local density $V_\text{eff}(\rho)$~\cite{tailleur2008statistical,cates2015motility}. Then, if fluctuations create a density excess  in a particular region, the particles there will move at a lower speed, implying that the incoming diffusive particle flux will be greater than the outgoing one, creating a mechanism for  instability. This model allows for a hydrodynamic-like description of the density and polarization fields, where it has been shown that in the limit of very large $\ell$ the density mode becomes unstable akin to spinodal decomposition if $-\derpar{V_\text{eff}}{\rho} > V_\text{eff}/\rho$. That is, if the velocity reduction is sufficiently drastic~\cite{tailleur2008statistical,cates2015motility}. For finite persistence lengths, corrections to this prediction appear, and the instability  develops only for $\ell$ greater than a threshold, in agreement with simulations~\cite{cates2015motility}.
Non-equilibrium thermodynamic formulations allow to obtain the binodal curves besides the spinodals~\cite{wittkowski2014scalar,cates2015motility,takatori2015towards,solon2018generalized,omar2023mechanical}.

Microscopically, the MIPS instability has been derived for lattice models, where it is possible to write the system dynamics in terms of a master equation~\cite{thompson2011lattice}. With the usual approximation of no correlations, MIPS is predicted to occur for analogous conditions as for ABP. 
For ABP, the MIPS explanation is realized by noting that hard-core collisions cause particles to take longer to travel a distance, i.e., the effective velocity is reduced by collisions. In Ref.~\cite{de2019active}, for large spatial dimensions, $d\gg1$, a kinetic-theoretic analysis allowed 
to compute the effective velocity reduction, obtaining $V_\text{eff}=V(1-\rho/\rho_\text{cr})$, where $\rho_\text{cr}$ is a characteristic density that depends on $d$. Also, using a mean-field approximation, the same dependence for $V_\text{eff}$ was found for ABP, although no analytical derivation of $\rho_\text{cr}$ was made~\cite{bialke2013microscopic}.
In Ref.~\cite{hancock2017statistical}, hydrodynamic equations showing MIPS were derived from a  mean-field kinetic theory for inertial  ABP.

Despite its importance, a complete microscopic derivation of MIPS for ABP has not yet been obtained.
Here, we present a kinetic theory description of ABP in the large persistence regime, from which we derive the conditions for MIPS to occur with a clear and identifiable mechanism for the reduction of the effective velocity. Kinetic theory is a powerful tool to coarse-grain microscopic models to obtain macroscopic equations for a reduced number of relevant fields (hydrodynamic-like equations)~\cite{soto2016kinetic}. 
In the case of active matter, kinetic equations have been successfully used in the low density limit for active particles presenting short-range aligning interactions~\cite{aranson2005pattern,bertin2006boltzmann,baskaran2008hydrodynamics,bertin2009hydrodynamic,ihle2014towards,mahault2018self}.
For microswimmers moving in a fluid  the interactions are mediated by the fluid and become long-range. In this case, a mean-field approximation, analogous to that used in plasma physics, has been used to study the instabilities that appear in these suspensions~\cite{saintillan2008instabilities,ezhilan2013instabilities} and the effects of fluctuations~\cite{qian2017stochastic}. Also, a kinetic analysis of the interactions of the swimmers with the fluid and among themselves has been used to characterize the rheology of bacterial suspensions~\cite{saintillan2010dilute,guzman2019nonideal}.
The case of ABP is more challenging for the construction of a kinetic theory, because the interactions are short ranged, but due to persistence, particles remain in contact for finite time and the usual concept of a collision is difficult to visualize. Here, however, we show that  it is indeed possible to formulate a kinetic description of ABP at moderate densities and large $\ell$, with collision events having well-defined pre and postcollisional states. The kinetic equation will be presented in general for $d$ spatial dimensions, but the explicit calculations will be performed in two dimensions. Finally, we want to emphasize that
ABP  have become a prototype for active matter because of its theoretical simplicity, the possibility to perform efficient simulations, and because, besides showing clustering, this model accurately describes the properties of many experimental realizations of non-inertial active matter such as Janus colloids~\cite{paxton2004catalytic,buttinoni2013dynamical,bialke2015active,ginot2015nonequilibrium}, Quincke rollers~\cite{bricard2013emergence,geyer2019freezing}, or active droplets~\cite{ramos2020bacteria}, just name just a few. 
The construction of a kinetic theory for ABP has, therefore, the possibility to help in the theoretical description of different phenomena shown by active matter.

\begin{figure}[htb]
\includegraphics[width=.9\columnwidth]{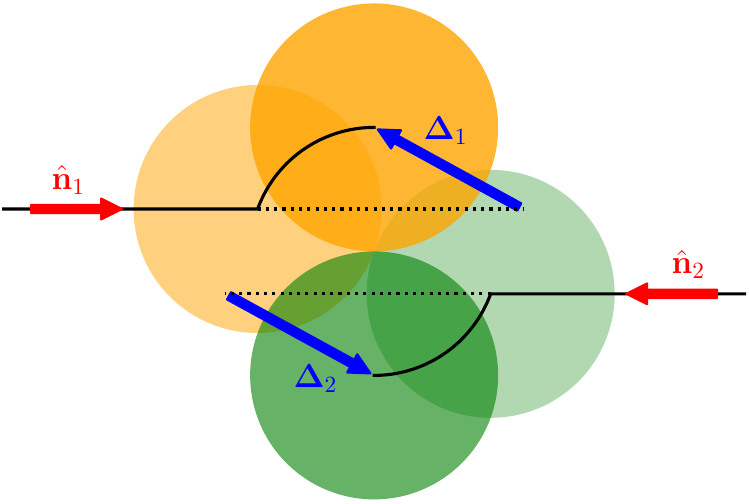}
\caption{Scheme of an effective collision for particles 1 (orange) and 2 (green). Directors $\HAT{n}_{1,2}$ are indicated by red arrows. The light colored circles show the initial state of the collision, while the dark colored circles show the state when the particles start to depart. The solid black lines show the actual trajectories up to the point of departure. The trajectories that the particles would had have followed without the collision are shown as black dotted lines. Finally, blue arrows show the effective displacements $\VEC{\Delta}_{1,2}$ caused by the collision . For simplicity, the figure only shows the case where $\HAT{n}_2=-\HAT{n}_1$, in which case the center of mass remains fixed.}
\label{fig.scheme}
\end{figure}

\textit{Effective collision theory for active Brownian particles.}
When two ABP meet, steric repulsion prevents them from continuing their free motion and they begin to slide in contact with each other. That is represented in Fig.~\ref{fig.scheme}: two particles  moving with velocity directors $\HAT{n}_1$ and $\HAT{n}_2$, get in contact at positions marked by light yellow and light green disks. They start to slide around each other until  they can detach again, indicated with dark yellow and dark green disks. Such  condition is mathematically satisfied when $(\HAT{n}_1-\HAT{n}_2)\cdot(\VEC{r}_1-\VEC{r}_2)=0$, being $\VEC{r}_i$ the position of the particle $i=1,2$. Trajectories displayed by solid black lines, while dotted lines would be the trajectories without the collision.
%  and $\HAT{n}_i$ its director, such that the particle velocity is $\VEC{V}_i=V\HAT{n}_i$.
The duration of this collision process $t_\text{col}$ is of the order of $\sigma/V$. Then, in the regime of large persistence lengths, $\ell\gg1$, the directors have hardly changed, allowing us to make the approximation that the directors remain constant in the process. 
What  changes are the particle positions. 
If we call $\VEC{r}_i^\text{ini}$ the particle positions when they first meet, $\VEC{r}_i^\text{end}$ the positions when they depart, and $\VEC\Delta_i^0=V\HAT{n}_i t_\text{col}$, the  distance travelled if the collision had not occurred, a collision results in net displacements $\VEC\Delta_i = \VEC{r}_i^\text{end}-\VEC{r}_i^\text{ini}-\VEC\Delta_i^0$.
Then, a collision can be modeled as an instantaneous process where the directors do not change, but the positions change as $\VEC{r}_i \rightarrow \VEC{r}_i + \VEC{\Delta}_i$ (depicted in Fig.~\ref{fig.scheme}).

It is possible to find an explicit expression for the displacements in terms of the particle directors and the unit vector $\hat{\bm{\sigma}}$ pointing from particle 1 to 2 at the beginning of the collision.  The calculation (see the Supplemental Material \cite{SuppMat}) consists of solving the equation of motion of the two particles with an additional normal force to maintain the impenetrability condition. The results are that $t^\text{col}= \frac{\sigma}{V|\HAT{n}_2-\HAT{n}_1|} \log|\tan(\theta/2)|$, with $\theta$ the angle between $\hat{\bm{\sigma}}$ and $\HAT{n}_2-\HAT{n}_1$, and
\begin{align}
\VEC{\Delta}_1 &= -\VEC{\Delta}_2 = -\sigma\frac{\hat{\bm{\sigma}}_\text{end}-\hat{\bm{\sigma}}}{2}
- V\Delta t^\text{col} \frac{\HAT{n}_1-\HAT{n}_2}{2},\\
&= -\frac{\sigma}{2}\left[\hat{\bm{\sigma}}_\text{end}-\hat{\bm{\sigma}}
- \log|\tan(\theta_0/2)| \frac{\HAT{n}_1-\HAT{n}_2}{|\HAT{n}_2-\HAT{n}_1|}\right].
 \end{align}
Here, $\hat{\bm{\sigma}}_\text{end}$ is the unit vector from 1 to 2 at the end of the collision, which is in the same plane as $\hat{\bm{\sigma}}$ and $(\HAT{n}_2-\HAT{n}_1)$, and perpendicular to the latter.
Note that although the system does not obey Galilean invariance, the displacements for the colliding particles are reciprocal. The collision time $t^\text{col}$ diverges for  head-on collisions ($\theta=\pi$), but it is an integrable divergence, giving finite results for the relevant calculations below.

\textit{Average velocity reduction.}
For a tagged particle, the displacement has a component perpendicular to its director that contributes to diffusion and mixing. More importantly for the purpose of understanding MIPS, there is a component parallel to the director $\Delta_\parallel=\VEC{\Delta}_1\cdot\HAT{n}_1$, which we show below to be negative on average. Therefore, the effective particle velocity is reduced as a result of collisions.

Before proceeding to derive the full kinetic theory, we present some elements of the theory by computing in a homogeneous system the average parallel displacement rate due to collisions, $\langle d\Delta_\parallel/dt|_\text{coll}\rangle$. Let $i=1$  be the tagged particle. For the collisions with particle 2, we assume the molecular chaos hypothesis for the precollisional states, corrected 
 with the static pair correlation function at contact $\chi$, as in the Enskog theory for moderately dense gases. That is, the collision rate  for the two particles  is $\chi(\rho) f(\HAT n_1)f(\HAT n_2) |V\sigma^{d-1}(\HAT n_2-\HAT n_1)\cdot\hat{\bm{\sigma}}| \Theta[-(\HAT n_2-\HAT n_1)\cdot\hat{\bm{\sigma}}]$, where $f$ is the distribution function.
The factor in absolute value represents the collision rate, which  is proportional to the  velocity $V$ multiplied by the effective cross section. Finally,  $\Theta$ is the Heaviside step function to select particles that are approaching \cite{soto2016kinetic}.
Assuming an equilibrium distribution in two dimensions, $f(\HAT n)=\rho/(2\pi)$, after integrating over all directions of $\HAT{n}_2$ and $\hat{\bm{\sigma}}$, we obtain $\langle d\Delta_\parallel/dt|_\text{coll}\rangle =  - \rho\chi \pi \sigma^2  V/4$ (see the Supplemental Material), which, as anticipated, is negative, indicating that 
collisions reduce the effective velocity of a particle to
\begin{align}
V_\text{eff} = V(1-\rho\chi \pi \sigma^2 /4) \label{eq.Veff}.
\end{align}

The spinodal density $\rho^*$ for the MIPS instability is given by the condition $-\derpar{V_\text{eff}}{\rho} = V_\text{eff}/\rho$~\cite{tailleur2008statistical,cates2015motility}, which upon substitution of Eq.~\eqref{eq.Veff} reads
%. The critical density, $\rho^*$, is the density value at the spinodal curve, that reads
% , which  for the critical density reads
\begin{align}
\rho^*\sigma^2 \left[\chi(\rho^*)+\rho^*\chi'(\rho^*)/2\right]=2/\pi, \label{eq.critical}
\end{align}
with $\chi'=d\chi/d\rho$, assuming that $\chi$ depends on the local density.
To evaluate Eq.~(\ref{eq.critical}), it is necessary to know the value of $\chi$, but it has not been determined for ABP~\footnote{Although the pair distribution function $g({\mathbf r})$ for ABPs has been measured in Refs.~\cite{de2018static,jeggle2020pair}, it is important to recall that for the kinetic equation, $\chi=\lim_{r\to\sigma^+}g({\mathbf r})$ must be evaluated only for precollisional states. Otherwise, artificially large values are obtained as reported in Ref.~\cite{soto2001statistical}.}. Therefore, we have to rely on expressions valid for elastic, passive, particles. 
The first approach can be to neglect correlations, $\chi=1$, approximation valid for very low densities. In this case, Eq.~\eqref{eq.critical} gives $\rho^*\sigma^2=2/\pi\approx0.64$, which is quite large, in the range of high densities and near close packing,  $\rho_\text{max}=2/(\sqrt{3}\sigma^2)$. Then, the assumption of no correlations is hard to justify. It is then necessary to use an expression for $\chi$ valid at moderate densities, such as that of Ref.~\cite{henderson1975simple}, for hard disks in equilibrium $\chi_\text{hd}=(1-7\pi \rho\sigma^2/64)/(1-\pi\rho\sigma^2/4)^2$. With this expression, 
the spinodal density is $\rho^*\sigma^2\approx0.32$ (area fraction $\phi^*=\pi\rho^*\sigma^2/4\approx0.25$), which is in the region of moderate densities where $\chi_\text{hd}$ is expected to be valid. The comparison with the simulation results for the spinodal curves is excellent. Simulations of ABP with hard disk interactions, Refs.~\cite{levis2017active,nie2020stability,digregorio2018full}, predict $\phi^*\approx 0.25$  for infinitely large $\ell$. Other authors carry out simulations for ABP interacting with softer potentials (see, e.g., Refs.~\cite{fily2014freezing,van2019interparticle,mallory2021dynamic}) predicting $\phi^*\approx0.30-0.35$. 
Softer potentials delay the MIPS transition, that is, the spinodal  line moves to higher densities~\cite{martin2021characterization,sanoria2021influence}. In both cases the agreement with our theory is excellent.

\textit{Kinetic theory.}
A kinetic theory that can be analyzed more formally to study MIPS, can be derived following the ideas presented above. In absence of collisions, the distribution function  evolves purely by the effects of free particle motion and rotational diffusion. Collisions can be included in the kinetic equation in a complete analogy to the Boltzmann--Enskog equation for moderately dense gases, except that instead of changing velocities, here each collision has the effect of displacing particles by an amount $\VEC\Delta_{i}$. Thus the equation for $f(\VEC r_1, \HAT n_1,t)$ reads
\begin{equation}\label{eq:kin}
    \frac{\partial f}{\partial t} + V\mathbf{\HAT{n}}_1\cdot\derpar{}{\VEC r_1} f = D_\text{r}\nabla^2_{\HAT{n}_1} f + J[f].
\end{equation}
The first three terms, up to the Laplace--Beltrami operator $\nabla^2_{\HAT{n}_1}$, are standard to account for the free streaming and rotational diffusion of the particles~\cite{bertin2006boltzmann,saintillan2008instabilities,saintillan2010dilute,romanczuk2012active,cates2013active,solon2015active,mahault2018self}. The  collisional term $J$  we propose is written, as in the Boltzmann--Enskog equation, as the difference of a gain and a loss term,
\begin{multline}\label{eq:operator}
    J[f] = \int \chi(\rho(\VEC r_1'+\VEC r_2')/2)    f(\VEC r_1',\HAT n_1) f(\VEC r_2',\HAT n_2)
 \\
\times |V\sigma^{d-1}(\HAT n_2-\HAT n_1)\cdot\hat{\bm{\sigma}}|\Theta[-(\HAT n_2-\HAT n_1)\cdot\hat{\bm{\sigma}}]
    \delta(\mathbf{r}_2^\prime-\mathbf{r}_1^\prime-\sigma\hat{\bm{\sigma}})\\
\times      [\delta(\mathbf{r}_1-\mathbf{r}_1^\prime-\VEC \Delta_1) -\delta(\mathbf{r}_1-\mathbf{r}_1^\prime)]d\mathbf{r}_1^\prime d\mathbf{r}_2^\prime d\mathbf{\hat{n}}_2d\hat{\bm{\sigma}}.
\end{multline}
The loss term, with the factor $\delta(\mathbf{r}_1-\mathbf{r}_1^\prime)$, indicates that a particle with position $\VEC r_1$ and director $\HAT n_1$ collides with a partner at the previously given rate, resulting in a decrease of $f(\VEC r_1, \HAT n_1,t)$. The gain term, with the factor $\delta(\mathbf{r}_1-\mathbf{r}_1^\prime-\VEC \Delta_1)$ accounts for the increase in $f(\VEC r_1, \HAT n_1,t)$ due to a particle located at $\mathbf{r}_1-\VEC \Delta_1$  colliding with a partner such that after the collision it ends at $\mathbf{r}_1$ with director $\HAT{n}_1$. 
 Both collision terms have the factor $\chi$ evaluated at the middle position of the two colliding particles.

The subtraction of the two Dirac deltas in Eq.\ (\ref{eq:operator}) represents the instantaneous particle teleportation at collisions, concept that is at the basis of the effective collision theory presented here. It is for the mass, the equivalent of the collisional transfer of momentum and energy for hard sphere systems, where these quantities are instantaneously exchanged between particles in a collision. As noted by Irving and Kirkwood, collisional transfers imply that momentum and energy are not locally conserved. 
However, by assuming that the momentum and energy flow along the line connecting the particle centers,  it is possible to define local stress tensors and heat fluxes
\cite{irving1950statistical,soto2016kinetic}.
Here we proceed analogously. For that, we note that we can write $\delta(\mathbf{r}-\mathbf{r}_a)-\delta(\mathbf{r}-\mathbf{r}_b) =-\nabla_\alpha\int_{\mathbf{r}_a}^{\mathbf{r}_b}\delta(\mathbf{r}-\mathbf{s})ds_\alpha$, where summation over repeated indices is used. 
With this expression, the collision term can be written as a divergence of a vector field, denoted by $\VEC G(\VEC r, \HAT n, t)$, with the form,
\begin{multline}\label{eq:operatorfull}
    J[f] = -\nabla_\alpha G_\alpha(\VEC r_1,\HAT n_1,t)
    =-\nabla_\alpha\bigg[\int \chi(\rho(\VEC r_1'+\VEC r_2')/2)\\
  \times   f(\VEC r_1',\HAT n_1) f(\VEC r_2',\HAT n_2) 
 |V\sigma^{d-1}(\HAT n_2-\HAT n_1)\cdot\hat{\bm{\sigma}}| \Theta[-(\HAT n_2-\HAT n_1)\cdot\hat{\bm{\sigma}}]\\
\times \delta(\mathbf{r}_2^\prime-\mathbf{r}_1^\prime-\sigma\hat{\bm{\sigma}})
  \int_{\mathbf{r}_1^\prime}^{\mathbf{r}_1^\prime+\VEC{\Delta_1}}\delta(\mathbf{r}_1-\mathbf{s})ds_\alpha d\mathbf{r}_1^\prime d\mathbf{r}_2^\prime d\mathbf{\hat{n}}_2d\hat{\bm{\sigma}} \bigg].
\end{multline}
With this expression,  integrating Eq.~\eqref{eq:kin} over $\HAT{n}_1$ gives the mass conservation equation 
\begin{align}
\frac{\partial \rho}{\partial t} = - \nabla\cdot\VEC J,
\end{align}
where 
\begin{align}
\rho(\VEC r,t)&=\int f(\VEC r, \HAT n, t) \, d\HAT n,\\
\VEC J(\VEC r,t)&=V\int f(\VEC r, \HAT n, t) \HAT n\, d\HAT n
+\int \VEC G(\VEC r, \HAT n, t) \, d\HAT n
\end{align}
are the density and the mass flux vector, respectively.

\begin{figure}[htb]
\includegraphics[width=.45\columnwidth]{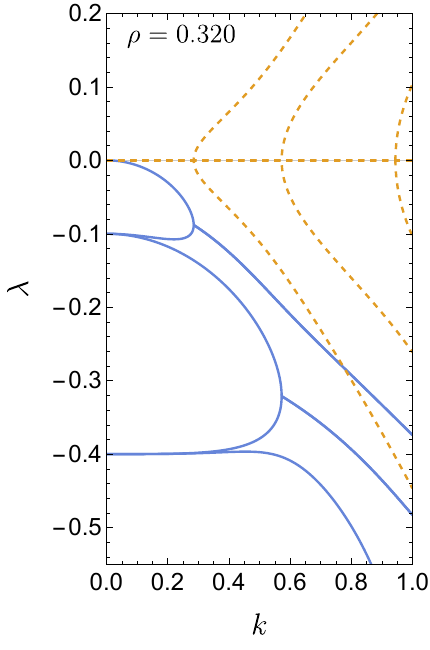}
\includegraphics[width=.45\columnwidth]{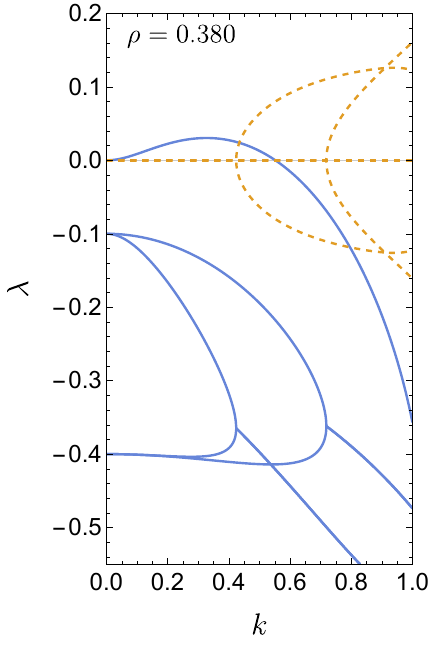}
\caption{Dynamical eigenvalues of the first 5 modes as a function of the wavevector $k$ for $\Dr=0.1$ ($\ell=10$), $\chi=\chi_\text{hd}$, and $\rho=0.32$ (left) and $\rho=0.38$ (right), obtained by truncating the dynamical matrix  to $p=-3,-2,\dots,3$. 
The real (imaginary) parts are shown with solid blue  (dashed orange) lines. 
Units have been chosen so that $V=\sigma=1$.}
\label{fig.ev}
\end{figure}

\textit{Linear perturbation and MIPS.}
Having derived the kinetic equation for ABP, we now proceed to study the stability of the homogeneous state to determine if MIPS is well described by this theory. For simplicity, we consider the two-dimensional case. First, it is easy to verify by direct substitution that the homogeneous and isotropic state, described by $f_0=\rho/(2\pi)$, is a stationary solution of the kinetic equation. Since the kinetic equation is homogeneous in space, we can use  spatial Fourier modes for the linear stability analysis. For the $\hat{\mathbf n}$-part of $f$, we consider a series of angular Fourier modes for the distribution of the director. In summary, we study solutions of the form
\begin{align}
f(\VEC r,\HAT n,t) = f_0 + e^{ i\VEC k\cdot\VEC r+\lambda t} \sum_m g_m e^{i m\phi } ,
\label{eq.linpert}
\end{align}
where $\lambda$ is the rate of amplification ($\Real\lambda>0$) or decay ($\Real\lambda<0$) of the perturbation. 
Projecting back the kinetic equation (\ref{eq:kin}) in the mode $e^{-ip\phi}$ and choosing $\VEC k =k \HAT{x}$, gives the eigenvalue problem for $\lambda$, 
\begin{align}
ik V \sum_{m} I_{pm}(k) g_m -\Dr p^2 g_p -\frac{ikV}{2} (g_{p+1}+g_{p-1}) = \lambda g_p, \label{eq.eigenvalue}
\end{align}
where the matrix elements $I_{pm}(k)$ are given in terms of the displacement $\VEC\Delta_1$ (see the Supplemental Material), and the prefactor $ik$ has been explicitly put  to reflect the effect of the divergence operator in the collision operator [Eq.~\eqref{eq:operatorfull}]. 
The eigenvalues  $\lambda_n$ can be obtained with increasing number of angular Fourier modes. Figure~\ref{fig.ev} shows two cases, one that is stable and one where the real part of an eigenvalue is positive, signaling the appearance of an instability, where the matrices have been truncated to 7 modes ($p=-3,-2,\dots,3$).
For any number of modes, it is found that for $k=0$ the eigenvalues are simply $\lambda_n=-\Dr n^2$, meaning that all modes are stable except for one that is marginal, the density mode. For finite but small wavevectors, the real part of the density mode eigenvalue is quadratic in $k$. 
Then, for the purpose of this letter, which is to show that  MIPS  is  predicted by kinetic theory, it is sufficient to show that for small wavevectors the density mode eigenvalue can be positive, analysis that can be done using perturbation theory. For that, a small $k$ expansion of the matrix elements is needed, which can be done analytically using the explicit expression of $\VEC{\Delta}_1$ (see the Supplemental Material). It is found  that for $I_{pm}(k=0)$ the only nonzero elements are when $m=p\pm1$, with $I_{\pm1,0}=\rho(2\chi+\rho\chi')\pi\sigma^2/8$, $I_{0,\pm1}=0$, and $I_{p,p\pm1}=\rho\chi\pi\sigma^2/8$ for the rest. Also needed is $dI_{00}(k=0)/dk=i2\sigma G(2\rho\chi + \rho^2\chi'  )/\pi$, where $G\approx 0.916$ is the Catalan constant. With these elements, perturbation theory gives 
$\lambda_0 = \frac{V^2}{2\Dr} \left[\frac{\rho\pi\sigma^2}{4} (2\chi + \rho\chi'  )\left(1-16 G \Dr\sigma/V\pi^2\right)-1 \right]k^2+{\cal O}(k^3)$. The spinodal density $\rho^*$ for MIPS is determined by the change of sign of the $k^2$ coefficient, resulting in a value that grows with $\Dr$. In the limit of large persistence lengths ($\Dr\to0$), where the present theory is valid, $\rho^*$ is obtained from the reduced equation \eqref{eq.critical}. Notably, the spinodal density obtained from the heuristic analysis of the effective velocity reduction coincides with that obtained from the formal analysis of the kinetic equation.

%\comRS{Este párrafo puede irse, ya no es necesario: For densities larger than the critical one, the amplification rate of the modes grows with $k$, without bound, which is physically impossible. This is not an artifact of the kinetic equation, but of the evaluation of $I_{pm}$ in the low $k$ limit} 
%\comRS{Deja esto, de una forma:} For the purpose of this letter, which is to show that  MIPS  is  predicted by kinetic theory, it is sufficient to show that for small wavevectors the eigenvalues can be positive.

%For $k=0$, the matrix elements can be computed analytically and the only nonzero elements are when $m=p\pm1$, with $I_{\pm1,0}=\rho(2\chi+\rho\chi')\pi\sigma^2/8$, $I_{0,\pm1}=0$, and $I_{p,p\pm1}=\rho\chi\pi\sigma^2/8$ for the rest.
%\comRS{Quizás para ganar espacio, lo que sigue se puede eliminar: This band-diagonal structure of the matrix allows to combine the effect of collisions with the streaming part in Eq.~\eqref{eq.eigenvalue}, but although suggestive, the $p$-dependence of the matrix elements does not allow to identify the effect of collisions as a mere reduction of the velocity. }

\textit{Discussion.}
The kinetic theory presented here is expected to be a valid formalism for different regimes occurring in active Brownian particles, and when additional interactions with external fields or between particles are considered. The only limitation is that the persistence length is large and that no long-lived bound states are formed, as happens for example in some non-reciprocal interactions~\cite{soto2014self,meredith2020predator}. As usual in kinetic theory, it is necessary to assume absence of all or at least some correlations in the precollisional state. Here, we were able to build the theory assuming that there are no director-director correlations, but that there are position correlations, which were considered in the factor $\chi$. To make more quantitative predictions, it is crucial to evaluate this factor. Note that no assumption has been made about the postcollisional states, which are indeed highly correlated.

The effective collision theory and the associated collision operator were obtained in the limit of infinite persistence lengths. 
For a more complete theory, it becomes relevant to develop a systematic approach to derive corrections for large but finite persistence lengths, where, as an effect of rotational diffusion, colliding particles can escape at different angles and with new directors, as in the case of tumbling particles in an array of fixed obstacles~\cite{saintillan2023dispersion}.  Formal methods like those used in Refs.~\cite{ihle2023asymptotically,ihle2023scattering} can be fruitful for this purpose.  
Heuristically, nevertheless, it is possible to advance that on average colliding particles will escape earlier for finite persistence lengths. This results in a less pronounced reduction of the effective velocity \eqref{eq.Veff}, implying that the spinodal density $\rho^*$ should grow with $\Dr$, consistently with experiments and simulations. This dependence of $\rho^*$ with $\Dr$ of kinematic origin should be added to the dependence found above in the linear stability analysis.

The application of the kinetic theory to an initially homogeneous gas correctly predicts MIPS without any ad-hoc hypothesis about the effective velocity. Rather, its reduction by collisions appears naturally and the predicted spinodal density shows an excellent agreement with the extrapolation of simulations to very large persistence lengths. 
Finally, for a complete analysis of the phase diagram, with the binodal curves besides the spinodal ones, it would be necessary to solve the stationary long-time non-linear dynamics of the kinetic equation, which will be the purpose of future work.

\acknowledgments
\textit{Acknowledgements}. This research was supported by the Fondecyt Grant No.~1220536 and Millennium Science Initiative Program NCN19\_170D of ANID, Chile. The work of R.B. is supported from Grant Number PID2020- 113455GB-I00. 
R.S.\ thanks Aparna Baskaran and Andrea Puglisi for inspiring discussions.

%\bibliography{biblio-KT-ABP}

%%%%%%%%%%%%%%%

\newpage
\appendix
\setcounter{secnumdepth}{2}

%%%%%%%%%%%%%%%%%%%%%%%%
%%%%%%%%%%%%%%%%%%%%%%%%
%%%%%%%%%%%%%%%%%%%%%%%%
\section{SUPPLEMENTAL MATERIAL}

\subsection{Explicit expressions for the displacements}

Consider two ABP particles with directors $\HAT{n}_{1,2}$ that enter in contact with a relative vector $\VEC r\equiv\VEC{r}_2-\VEC{r}_1=\sigma\hat{\bm{\sigma}}_0$. For clarity, in this Supplementary Material, we will call $\hat{\bm{\sigma}}_0$ the initial relative unit vector and $\hat{\bm{\sigma}}$ the instantaneous relative unit vector during the collision. The relative velocity prior to the collision is $V(\HAT{n}_2-\HAT{n}_1)$. Then, the condition of collision is that the two particles are approaching, that is, $(\HAT{n}_2-\HAT{n}_1)\cdot\hat{\bm{\sigma}}_0<0$.
We define the unit vector $\HAT{z}$ as parallel to $(\HAT{n}_2-\HAT{n}_1)$, that is  $\HAT{z}=(\HAT{n}_2-\HAT{n}_1)/|\HAT{n}_2-\HAT{n}_1|=(\HAT{n}_2-\HAT{n}_1)/\sqrt{2-2\HAT{n}_1\cdot\HAT{n}_2}$, and we define $\theta_0$ as the  angle between $\hat{\bm{\sigma}}_0$ and $\HAT{z}$, that is, $\cos\theta_0=\hat{\bm{\sigma}}_0\cdot\HAT{z}$. Note that the collision condition implies that $\cos\theta_0< 0$. Similarly, we define the instantaneous angle $\theta$ such that $\cos\theta=\hat{\bm{\sigma}}\cdot\HAT{z}$

During the collision, the equations of motion of the two particles are
\begin{align}
\dot{\VEC{r}}_1 &= V \HAT{n}_1 - N\hat{\bm{\sigma}} \label{eq.r1coll},\\
\dot{\VEC{r}}_2 &= V \HAT{n}_2 + N\hat{\bm{\sigma}}\label{eq.r2coll},
\end{align}
where $N$ is the normal that guarantees that they do not penetrate, which is computed imposing that $(\dot{\VEC{r}}_2 -\dot{\VEC{r}}_1)\cdot\hat{\bm{\sigma}}=0$, resulting in $N=V(\HAT{n}_1- \HAT{n}_2)\cdot\hat{\bm{\sigma}}/2$. The normal vanishes for $\theta=\pi/2$, which corresponds  to the moment of detachment of the two particles.

Adding \eqref{eq.r1coll} and \eqref{eq.r2coll}, gives for the center of mass $\VEC{R}\equiv (\VEC{r}_1+\VEC{r}_2)/2$,
\begin{align}
\dot{\VEC{R}} &= V (\HAT{n}_1 + \HAT{n}_2)/2. \label{eq.Rcoll}
\end{align}

For the relative vector, as the radial component is fixed by the normal, the only relevant components are the tangential ones. We define the unit vector $\hat{\bm{\theta}}$ to be in the same plane as $\HAT{z}$ and $\hat{\bm{\sigma}}$, but perpendicular to $\hat{\bm{\sigma}}$. We complete the set of spherical unit vectors with $\hat{\bm{\phi}}=\hat{\bm{\sigma}}\times\hat{\bm{\theta}}$. Then, the tangential components for $\dot{\VEC{r}}=\dot{\VEC{r}}_2-\dot{\VEC{r}}_1$ are
\begin{align}
\dot{\VEC{r}}\cdot\hat{\bm{\phi}} &= V(\HAT{n}_2-\HAT{n}_1)\cdot\hat{\bm{\phi}}\\
&= V|\HAT{n}_2-\HAT{n}_1|\HAT{z}\cdot\hat{\bm{\phi}}\\
&=0
\end{align} 
and 
\begin{align}
\dot{\VEC{r}}\cdot\hat{\bm{\theta}} &= V(\HAT{n}_2-\HAT{n}_1)\cdot\hat{\bm{\theta}} \\
&= V|\HAT{n}_2-\HAT{n}_1|\HAT{z}\cdot\hat{\bm{\theta}}, \\
\sigma\dot\theta&=-V|\HAT{n}_2-\HAT{n}_1|\sin\theta.
\end{align} 
That is, the relative unit vector remains in the $\HAT{z}$--$\hat{\bm{\sigma}}_0$ plane, with the angle evolving according to
\begin{align}
\dot\theta&=-V|\HAT{n}_2-\HAT{n}_1|\sin\theta/\sigma.
\end{align} 
This differential equation must be integrated from the initial condition $\theta_0$ to the moment of detachment at $\pi/2$. That gives the total collision time
\begin{align}
\Delta t^\text{col} &= -\frac{\sigma}{V|\HAT{n}_2-\HAT{n}_1|} \int_{\theta_0}^{\pi/2} \frac{d\theta}{\sin\theta}\\
&= \frac{\sigma}{V|\HAT{n}_2-\HAT{n}_1|} \log\tan(\theta_0/2). \label{sm.dtcol}
\end{align}

At the end of the collision the relative vector, characterized by having $\theta=\pi/2$ is given by \begin{align}
\hat{\bm{\sigma}}_\text{end} &= \frac{\hat{\bm{\sigma}}_0 - (\hat{\bm{\sigma}}_0\cdot\HAT{z})\HAT{z}}
{|\hat{\bm{\sigma}}_0 - (\hat{\bm{\sigma}}_0\cdot\HAT{z})\HAT{z}|}\\
&=\frac{\hat{\bm{\sigma}}_0 - \cos\theta_0\HAT{z}}{\sin\theta_0}.
\end{align}
Hence, during the collision, the relative vector has experienced a total displacement
\begin{align}
\Delta\VEC{r}^\text{col} = \sigma(\hat{\bm{\sigma}}_\text{end} -\hat{\bm{\sigma}}_0).
\end{align}
At the same time, during the collision,  the center of mass has displaced
\begin{align}
\Delta\VEC{R}^\text{col} = V (\HAT{n}_1 + \HAT{n}_2)\Delta t^\text{col} /2,
\end{align}
where we used Eq.\ \eqref{eq.Rcoll}.
With this, we obtain the total travelled distance  by the particles during the collision
\begin{align}
\Delta\VEC{r}_1^\text{col} &= \Delta\VEC{R}^\text{col} - \Delta\VEC{r}^\text{col}/2,\\
\Delta\VEC{r}_2^\text{col} &= \Delta\VEC{R}^\text{col} + \Delta\VEC{r}^\text{col}/2.
\end{align}

Now, we are in condition to write down the effective displacements $\VEC{\Delta}_{i}$ as the total travelled distance during the collision $\Delta\VEC{r}_{i}^\text{col}$, minus  what they would have travelled during the same time as if there were have been no collision $V\HAT{n}_{i}\Delta t^\text{col}$, which simplify to
\begin{align}
\VEC{\Delta}_1 &= -\sigma\frac{\hat{\bm{\sigma}}_\text{end}-\hat{\bm{\sigma}}_0}{2}
- V\Delta t^\text{col} \frac{\HAT{n}_1-\HAT{n}_2}{2},\\
\VEC{\Delta}_2 &= -\VEC{\Delta}_1.
\end{align}
Recalling the expressions for $\Delta t^\text{col}$, $\hat{\bm{\sigma}}_\text{end}$, $\HAT{z}$, and $\theta_0$, we notice that effective displacements depend only on $\HAT{n}_2-\HAT{n}_1$ and $\hat{\bm{\sigma}}_0\cdot(\HAT{n}_2-\HAT{n}_1)$. We also note that, by using \eqref{sm.dtcol}, the effective displacements do not depend on $V$ and are proportional to the diameter $\sigma$.

%%%%%%%%%%%%%%%%%%%%%%%%
%%%%%%%%%%%%%%%%%%%%%%%%
%%%%%%%%%%%%%%%%%%%%%%%%

\subsection{Parametrization in two spatial dimensions}

In three dimensions, the instantaneous unit relative vector $\hat{\bm{\sigma}}$ is parametrized by $\theta\in[0,\pi]$, and the azimuthal angle $\varphi\in[0,2\pi]$. In two dimensions, there is no azimuthal angle and now $\theta$ lies in the full range $[0,2\pi]$. Then, two situations can can take place: if $\pi/2\leq\theta_0<\pi$, at the end of the collision $\theta_\text{end}=\pi/2$, but if $\pi<\theta_0\leq3\pi/2$, at the end of the collision $\theta_\text{end}=3\pi/2$. Consistently, to be valid in all cases in 2D,  Eq.~\eqref{sm.dtcol} has to be modified to  
\begin{align}
\Delta t^\text{col}= \frac{\sigma}{V|\HAT{n}_2-\HAT{n}_1|} \log|\tan(\theta_0/2)|.
\end{align}

Writing the directors as $\HAT{n}_i=(\cos\phi_i,\sin\phi_i)$ and defining the rotation matrix
\begin{align}
R(\alpha)=\begin{pmatrix}
\cos\alpha & -\sin\alpha\\
\sin\alpha & \cos\alpha
\end{pmatrix},
\end{align}
we have
\begin{align}
\HAT{z} &= \left(\frac{\cos\phi_2-\cos\phi_1}{\sqrt{2-2\cos(\phi_2-\phi_1)}}, \frac{\sin\phi_2-\sin\phi_1}{\sqrt{2-2\cos(\phi_2-\phi_1)}}\right),\\
\hat{\bm{\sigma}}_0&=R(-\theta) \HAT{z},\\
\intertext{and}
\hat{\bm{\sigma}}_\text{end}&=\begin{cases}
R(-\pi/2) \HAT{z}, & \pi/2\leq\theta_0<\pi\\
R(\pi/2) \HAT{z}, & \pi<\theta_0\leq3\pi/2.
\end{cases}
\end{align}

\begin{widetext}
With this parametrization in terms of $\phi_1$, $\phi_2$, and $\theta$ (now, the angle at the beginning of the collision), the different expressions reduce to
\begin{gather}
|\HAT{n}_2-\HAT{n}_1| = \sqrt{2 - 2 \cos(\phi_1 - \phi_2)},\\
(\HAT n_2-\HAT n_1)\cdot\hat{\bm{\sigma}}= \sqrt{2 - 2 \cos(\phi_1 - \phi_2)}\cos\theta,\\
    |(\HAT n_2-\HAT n_1)\cdot\hat{\bm{\sigma}} | \bm{\Delta}_1=\frac{\sigma|\cos\theta|}{2} \bigg\{\big[\cos(\phi_2 - \theta) - 
   \cos(\phi_1 - \theta) + (\cos\phi_2 - \cos\phi_1) \log
     |\tan(\theta/2)| \pm (\sin\phi_1 - \sin\phi_2)\big]\HAT{x}\nonumber\\   
+ \big[\sin(\phi_2 - \theta) - 
   \sin(\phi_1 - \theta) + (\sin\phi_2 - \sin\phi_1) \log
     |\tan(\theta/2)| \pm (\cos\phi_2 - \cos\phi_1)\big]\HAT{y}\bigg\},\label{SM.Deltaexplicit}
\end{gather}
where the positive sign is for $\pi/2<\theta<\pi$ and the negative one for $\pi<\theta<3\pi/2$.
\end{widetext}

%%%%%%%%%%%%%%%%%%%%%%%%
%%%%%%%%%%%%%%%%%%%%%%%%
%%%%%%%%%%%%%%%%%%%%%%%%

\subsection{Linear stability analysis}

\subsubsection{Expansion in Fourier modes}
We consider the expansion
\begin{align}
f(\VEC r,\HAT n,t) = f_0 + f_1(\VEC r,\HAT n,t),
\label{eq.linpert1SM}
\end{align}
where $f_0=\rho/(2\pi)$ is the reference isotropic and homogeneous distribution in 2D, and the perturbation is
\begin{align}
 f_1(\VEC r,\HAT n,t)=
e^{ i\VEC k\cdot\VEC r+\lambda t} \sum_m g_m e^{i m\phi }.
\label{eq.linpert2SM}
\end{align}

\subsubsection{Eigenvalue problem}
Substituting this expansion, with $\VEC k=k\HAT{x}$, in the kinetic equation 
\begin{equation}\label{eq:kinSM}
    \frac{\partial f}{\partial t} + V\mathbf{\HAT{n}_1}\cdot\derpar{}{\VEC r_1} f = D_\text{r}\nabla^2_{\HAT{n}_1} f + J[f,f],    
\end{equation}
gives for the first three terms
\begin{align}\label{eq2SM}
e^{ i\VEC k\cdot\VEC r+\lambda t} \sum_m  g_m \left[\lambda + \frac{i kV}{2} \left(e^{i\phi_1}+e^{-i\phi_1}\right)-\Dr m^2 \right] e^{im\phi_1},
\end{align}
where we used that $\VEC k\cdot\HAT{n}_1=kn_{1x} = k\cos\phi_1$. Projecting into the $p$ Fourier mode, that is, computing $(2\pi)^{-1}\int_0^{2\pi} d\phi_1 e^{-ip\phi_1}$ over Eq.~\eqref{eq2SM} results in
\begin{align}\label{eq3SM}
\lambda g_p+ \frac{i kV}{2} \left(g_{p+1}+g_{p-1}\right)-\Dr p^2 g_p,
\end{align}
where we have factored out the exponential prefactor. 

To evaluate the contribution of collisions, we first note that, as usual in Enskog theories,  the collision operator is not  bilinear  because $\chi$ depends on the local density. Indeed,
\begin{align}
\chi(\VEC r) &= \chi(\rho(\VEC r))\\
&= \chi(\rho + e^{i \VEC k \cdot \VEC{r}} \rho_1 )\\
&\approx \chi(\rho) + \chi'(\rho) e^{i \VEC k \cdot \VEC{r}} \rho_1,
\end{align}
where we linearized for small perturbations, $\rho_1=2\pi g_0$ is the local correction of the density, and used the notation $\chi'=d\chi/d\rho$.

\begin{widetext}
With these elements, after integrating out all the Dirac delta functions of $J[f]$ in Eq.~(7) of the main text, one obtains to linear order
\begin{align}
J[f_0+f_1]=& \int 
\left\{\chi(\rho) f_0\left[f_1(\HAT n_1) +e^{i\VEC k\cdot\sigma\hat{\bm \sigma}} f_1(\HAT n_2)\right]
+\chi'(\rho) f_0^2  \rho_1 e^{i\VEC k\cdot\sigma\hat{\bm \sigma}/2}\right\}
\nonumber\\
&\times \left(e^{-i\VEC k\cdot\VEC \Delta_1}-1\right) |V\sigma^{d-1}(\HAT n_2-\HAT n_1)\cdot\hat{\bm{\sigma}}| \Theta[-(\HAT n_2-\HAT n_1)\cdot\hat{\bm{\sigma}}] d\mathbf{\hat{n}}_2d\hat{\bm{\sigma}}, \label{eqsm.linear1}
\end{align}
where we also factored out the exponential prefactor.

Replacing the expansion~\eqref{eq.linpert2SM} and using  the parametrization for $\HAT{n}_{i}$ and $\hat{\bm{\sigma}}$ given in the previous section, we project into the $p$ Fourier mode, that is we compute $(2\pi)^{-1}\int_0^{2\pi} d\phi_1 e^{-ip\phi_1} J[f_0+f_1]$, for $\VEC k=k\HAT{x}$:
\begin{align}
\frac{ikV \sigma}{(2\pi)^2} \int_0^{2\pi} d\phi_1 \int_0^{2\pi} d\phi_2 \int_{\pi/2}^{3\pi/2} d\theta 
 e^{-ip\phi_1}
 \left\{\chi(\rho) \rho \sum_m \left[ e^{im \phi_1} +e^{i\VEC k\cdot\sigma\hat{\bm \sigma}} e^{im\phi_2}\right] g_m
+\chi'(\rho) \rho^2  g_0 e^{i\VEC k\cdot\sigma\hat{\bm \sigma}/2}\right\}
\nonumber\\
\times \left(\frac{e^{-i\VEC k\cdot\VEC \Delta_1}-1}{ik}\right)
 |(\HAT n_2-\HAT n_1)\cdot\hat{\bm{\sigma}}| =ikV\sum_m I_{pm}(k) g_m, \label{SM.eqIpm}
\end{align}
which defines the matrix elements $I_{pm}(k)$. We used that the common factor $ \left(e^{-i\VEC k\cdot\VEC \Delta_1}-1\right)$ in Eq.~\eqref{eqsm.linear1} vanishes in the limit $k\to0$, consistent with the divergence operator in front of the collision term in Eq.~(8) of the main text, to put an explicit prefactor $k$ in Eq.~\eqref{SM.eqIpm}. With this choice, the matrix elements remain finite in the limit $k\to0$ and can be evaluated for any value of $k$ by direct numerical integration using the explicit form of $\VEC \Delta_1$.
\end{widetext}

Collecting \eqref{eq3SM} and \eqref{SM.eqIpm}, the eigenvalue problem for the growth rates $\lambda$ reads
\begin{align}
ik V \sum_{m} I_{pm}(k) g_m -\Dr p^2 g_p -\frac{ikV}{2} (g_{p+1}+g_{p-1}) &= \lambda g_p, \label{eqsm.ev1}
\end{align}
which can be written as
\begin{align}
\sum_m \Lambda_{pm}(k) g_m&= \lambda g_p. \label{SMeq.eigenvalue}
\end{align}

\subsubsection{Largest eigenvalue for small wavevectors}

The structure of the eigenvalue problem \eqref{eqsm.ev1} allows to analyze the eigenvalues perturbatively in powers of $k$. Specifically, we are interested in the largest eigenvalue, which can become positive, indicating the emergence of an instability. For that, we first expand the matrix $\Lambda$ in powers of $k$, $\Lambda=\Lambda^{(0)} + k \Lambda^{(1)} + k^2\Lambda^{(2)}+\dots$, where
\begin{align}
\Lambda^{(0)}_{pm} &=-\Dr p^2\delta_{pm}, \label{eqsm.L0}\\
\Lambda^{(1)}_{pm} &=iV I_{pm}(0) - \frac{iV}{2}\left(\delta_{p,m+1}+\delta_{p,m-1}\right), \label{eqsm.L1}\\
\Lambda^{(2)}_{pm} &= iV I'_{pm}(0), \label{eqsm.L2}
\end{align}   
with $I'_{pm}=dI_{pm}/dk$.

The eigenvalue problem then reads
\begin{multline}
(\Lambda^{(0)} + k \Lambda^{(1)} + k^2\Lambda^{(2)}+\dots)(g^{(0)}+kg^{(1)}+k^2g^{(2)}+\dots) 
=\\
(\lambda^{(0)} + k \lambda^{(1)} + k^2\lambda^{(2)}+\dots)(g^{(0)}+kg^{(1)}+k^2g^{(2)}+\dots) , \label{eqsm.evgeneral}
\end{multline}
which can be analyzed order by order. To order $k^0$, the equation reduces to
\begin{align}
\Lambda^{(0)}g^{(0)}=\lambda^{(0)}g^{(0)},
\end{align}
which has the simple solution $\lambda^{(0)}_n=-D_r n^2$, for $n\in\mathbb{Z}$. For $n=0$, the eigenvalue vanishes at $k=0$, associated to the mass conservation. All the others are strictly negative. By continuity, the $n=0$ eigenvalue is the only one that can become positive for small $k$. We will therefore study this eigenvalue to different orders in $k$. For simplicity, we will omit in what follows the subscript $n$. The corresponding eigenvector is simply
\begin{align}
g^{(0)}_p=\delta_{p,0} . \label{eqsm.vector0}
\end{align}

To order $k^1$, Eq.~\eqref{eqsm.evgeneral} reads
\begin{align}
\Lambda^{(0)}g^{(1)}=\lambda^{(1)}g^{(0)}-\Lambda^{(1)}g^{(0)}, \label{eqsm.k1}
\end{align}
where we used that $\lambda^{(0)}=0$. The hermitian matrix $\Lambda^{(0)}$ is singular. Hence, for this equation to have a solution, the right hand side must be orthogonal to its nullspace, $g^{(0)}$. That is, we must have
\begin{align}
\lambda^{(1)} g^{(0)T} g^{(0)}-g^{(0)T}\Lambda^{(1)}g^{(0)}=0, 
\end{align}
where $T$ means transpose. Below, we show that $\Lambda^{(1)}$ vanishes identically on the diagonal and the only non-zero elements are for $p=m\pm1$. This implies that
\begin{align}
\lambda^{(1)}=0
\end{align}
and that the solution of Eq.~\eqref{eqsm.k1} is
\begin{align}
g^{(1)}_p = \Lambda^{(1)}_{p0}(\delta_{p,1}+\delta_{p,-1})/\Dr. \label{eqsm.vector1}
\end{align}

Finally, to order $k^2$, Eq.~\eqref{eqsm.evgeneral} simplifies to
\begin{align}
\Lambda^{(0)}g^{(2)}=\lambda^{(2)}g^{(0)}-\Lambda^{(1)}g^{(1)}-\Lambda^{(2)}g^{(0)}.\label{eqsm.k2}
\end{align}
Again, for this equation to have a solution,  the right hand side must be orthogonal to $g^{(0)}$, implying that
\begin{align}
\lambda^{(2)}= g^{(0)T}\Lambda^{(1)}g^{(1)} + g^{(0)T}\Lambda^{(2)}g^{(0)},
\end{align}
where we used that $g^{(0)T} g^{(0)}=1$. 

In summary, the largest eigenvalue is
\begin{align}
\lambda= \left(\frac{\Lambda^{(1)}_{0,1}\Lambda^{(1)}_{1,0} + \Lambda^{(1)}_{-1,0}\Lambda^{(1)}_{0,-1}}{\Dr} + \Lambda^{(2)}_{0,0}\right) k^2+{\cal O}(k^3), \label{eqsm.lambda2a}
\end{align} 
where we used the explicit forms of $g^{(0)}$ and $g^{(1)}$ [Eqs.~\eqref{eqsm.vector0} and \eqref{eqsm.vector1}] to write the result in terms of the matrix elements. Hence, to determine if it is possible for $\lambda$ to become positive, we need to evaluate $\Lambda^{(1)}$ and the matrix element $\Lambda^{(2)}_{0,0}$.

\begin{widetext}

It is possible to obtain exact analytical expressions for the relevant terms in the matrix expansion. Indeed, taking the limit,
\begin{align}
I_{pm}(0)= -\frac{\sigma}{(2\pi)^2} \int_0^{2\pi} d\phi_1 \int_0^{2\pi} d\phi_2 \int_{\pi/2}^{3\pi/2} d\theta 
 e^{-ip\phi_1}
 \bigg[\rho\chi (e^{i m\phi_1 }+e^{i m\phi_2}) + \rho^2\chi' \delta_{m0} \bigg]
 |(\HAT n_2-\HAT n_1)\cdot\hat{\bm{\sigma}}| \Delta_{1x} .
\end{align}
With the explicit expression Eq.~\eqref{SM.Deltaexplicit}, it is direct to obtain that
\begin{align}
I_{p,p+1}(0) &= \frac{\rho\pi\sigma^2}{8}\times\begin{cases}
2\chi+\rho\chi' & \text{if\ }p=-1\\
0 &  \text{if\ }p=0\\
\chi & \text{otherwise}
\end{cases}, &
I_{p,p-1}(0) &= \frac{\rho\pi\sigma^2}{8}\times\begin{cases}
2\chi+\rho\chi' & \text{if\ }p=1\\
0 &  \text{if\ }p=0\\
\chi & \text{otherwise}
\end{cases},
\end{align}
while all other elements vanish. With the second contribution in Eq.~\eqref{eqsm.L1}, it results that, as advanced, $\Lambda^{(1)}$ is not vanishing only for $p=m\pm1$.

Finally, we also need $\Lambda^{(2)}_{0,0} = iV I'_{0,0}(0)$. It is given by
\begin{align}
\Lambda^{(2)}_{0,0}&= \frac{V\sigma}{2(2\pi)^2} (2\rho\chi + \rho^2\chi'  ) 
\int_0^{2\pi} d\phi_1 \int_0^{2\pi} d\phi_2 \int_{\pi/2}^{3\pi/2} d\theta  
 |(\HAT n_2-\HAT n_1)\cdot\hat{\bm{\sigma}}| \Delta_{1x} ( \sigma_{x}- \Delta_{1x})\\
 &= -2V\sigma G(2\rho\chi + \rho^2\chi'  )/\pi,
\end{align}
where $G\approx 0.916$ is the Catalan constant.

Substituting these results in Eq.~\eqref{eqsm.lambda2a} gives
\begin{align}
\lambda = \frac{V^2}{2\Dr} \left[\frac{\rho\pi\sigma^2}{4} (2\chi + \rho\chi'  )\left(1-16 G \Dr\sigma/V\pi^2\right)-1 \right]k^2+{\cal O}(k^3).
\end{align}
The eigenvalue becomes positive when the square bracket vanishes,
\begin{align}
\frac{\rho\pi\sigma^2}{4}(2\chi + \rho\chi'  )\left(1-16 G \Dr\sigma/V\pi^2\right)-1=0.
\end{align}
Its solution gives the spinodal transition density $\rho^*$. In the limit of large persistence ($\Dr\to0$), the equation for $\rho^*$ reduces to
\begin{align}
\frac{\rho\pi\sigma^2}{4}(2\chi + \rho\chi'  )-1=0,
\end{align}
which is the same to that found when analyzing the effective velocity reduction.

\end{widetext}

\end{document}